\documentclass{SPAICE}

\def\authorEmail{amit.reza@oeaw.ac.at}
\def\AuthorShort{A. Reza et al.}

\newcommand{\tpprof}{(T_{\text{gas}}, p_{\text{gas}})}
\newcommand{\Rsq}{\ensuremath{\text{R}^{2}}}
\author[1]{Amit Reza \thanks{Corresponding author. E-Mail: \authorEmail}}
\author[1]{Ludmila Carone}
\author[1,2]{Christiane Helling}
\affil[1]{Space Research Institute, Austrian Academy of Sciences, Schmiedlstrasse 6, A-8042 Graz, Austria}
\affil[2]{Institute for Theoretical Physics and Computational Physics, Graz University of Technology, Petersgasse 16 8010 Graz}

\title{Faithful Neural Embeddings for 3D Exoplanet Climate Modeling}
\begin{document}

\maketitle

\begin{abstract}
\noindent With the rapid advancement of telescopes like JWST and Ariel, there is an urgent need for efficient 3D climate models to interpret observations of exoplanet atmospheres. Traditional 3D general circulation models (GCMs) are computationally intensive, prompting the development of machine learning (ML) emulators to accelerate simulations. Recent work, such as that by Plaschzug et al. 2026 \cite{plaschzug2026accelerating}, uses a dense neural network (DNN) to predict local gas temperatures and winds from input parameters, including local gas pressure, spatial coordinates (longitude and latitude), and global temperature. However, this model relies on predicting individual temperature values (points) at specific grid points, which can be limited by the resolution and constraints of the training grid. In this work, we investigate a couple of alternative frameworks based on latent-space representations of local gas temperature ($\text{T}_{\text{gas}}$) to obtain a faithful, low-dimensional representation of these profiles. This represents the first step toward developing a latent space regression model, offering a structurally cohesive alternative to the existing point-wise prediction method \cite{plaschzug2026accelerating}. By capturing the optimal embedding space of atmospheric data, our proposed framework can produce simulated profiles while maintaining computational efficiency, making it suitable for large-scale exoplanet ensemble studies. 
\end{abstract}

\section{Introduction}
In the last few decades, exoplanet research has started shifting from discovery to the rigorous characterization of their climates and atmospheric profiles \cite{madhusudhan2019exoplanetary}. Many of these worlds exist in regimes far beyond the solar system standard \cite{seager2010exoplanet}, from the blistering irradiation of ``Hot Jupiters" \cite{fortney2008unified} to the permanent thermal gradients of tidally locked planets \cite{showman2011atmospheric}, driving complex, non-intuitive patterns of circulation and chemical behavior \cite{parmentier20183d}.

A key challenge in understanding exoplanet atmospheres is how cloud formation and non-equilibrium chemistry affect observable spectra \cite{Helling2006, Woitke2020}. These processes influence atmospheric opacity and critical spectral features for interpretation.
While these kinetic processes occur at microphysical scales, their global impact can be fully quantified only through 3D General Circulation Models (GCMs) \cite{plaschzug2026accelerating}. These models are essential for mapping how local cloud formation and chemical gradients interact with large-scale atmospheric dynamics. However, the high complexity of 3D GCMs, coupled with radiative transfer and fluid dynamics, makes them computationally expensive, often requiring weeks of high-performance computing (HPC) time for a single planetary scenario. To overcome this, Machine Learning (ML) has emerged as a vital acceleration tool. By training surrogate models or 3D GCM emulators, researchers can simulate complex atmospheric behavior at a fraction of the cost of traditional methods, enabling large-scale grid simulations necessary for modern spectroscopic retrieval.
Plaschzug et al. 2026 \cite{plaschzug2026accelerating} have successfully derived a dense neural network model that can emulate 3D GCM output. This work focuses on the development of an alternative method, called a ``latent-space regressor'', to develop the 3D GCM emulator. Our primary objective is to understand whether the proposed framework in this work provides a more structurally cohesive alternative to point-wise \cite{plaschzug2026accelerating} predictions. 
\section{Dataset}
\begin{figure}[ht!]
\centering
\includegraphics[width=.99\columnwidth]{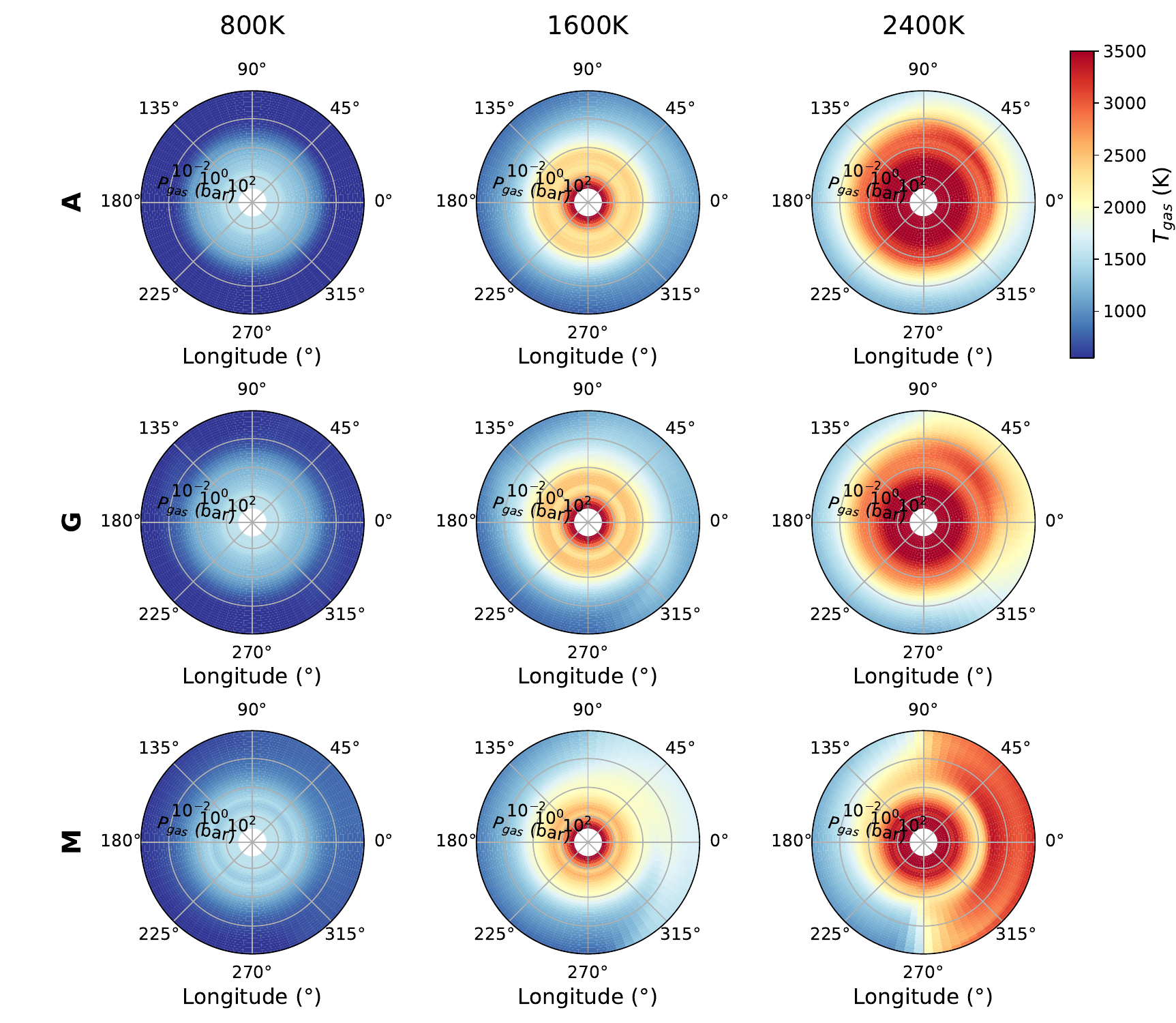}
\caption{Radial gas temperature maps for planets within the in-house 3D GCM framework. The panels illustrate equatorial slice plots of ($\text{T}_{\text{gas}}$, $\text{p}_{\text{gas}}$) across three global temperature regimes: 800 K (left), 1600 K (center), and 2400 K (right). Host star types vary by row, descending from A-type to G-type and M-type stars \cite{plaschzug2026accelerating} }
\label{fig:EQ_slice}
\end{figure}

For this study, we have used our in-house 3D GCM grid software package, called $\texttt{IWF}-\texttt{ExoRad}$ \cite{plaschzug2026accelerating}, to obtain climate simulations. The data grid covers 60 planets, each orbiting one of five different types of stars. The planets have global average temperatures ranging from a cool 400 K to a scorching 2600 K. Each simulation maps 42 pressure layers, 45 latitude points, and 72 longitude points. Each point in the grid is defined by its specific location and environment (star type, global temperature, local gas pressure, latitude, and longitude). At each of these points, we track four key results: (a) Local gas temperature $\text{T}_{\text{gas}}$, (b) Zonal wind (East-West speed), (c) Meridional Wind (North-South speed), and (d) Vertical Wind (Up-Down speed). However, this work primarily investigates the scope of predicting local gas temperature at arbitrary grid points. 
Figure \ref{fig:EQ_slice} shows an example of equatorial slice plots of ($\text{T}_{\text{gas}}$, $\text{p}_{\text{gas}}$). 
\section{Method}
This work focuses on developing two different ML frameworks, defined as (a) \textbf{Latent Space Regressor} (LSR) framework using a combination of autoencoder and dense neural network, and (b) \textbf{Generative Predictor} framework using Conditional Variational Autoencoder (CVAE) \cite{sohn2015learning}. Mathematically, we define $(\text{T}_{\text{gas}}$, $\text{p}_{\text{gas}})$ as a vector $\mathbf{y} \in \mathbb{R}^{n}$, where n is 47 samples, each of these samples represents a discrete pair of local gas-pressure and temperature. The parameters $(\theta, \phi, \text{T}_{\text{global}}, \text{T}_{\text{eff}})$ are considered as an input vector $\mathbf{x}$. Hence, in our formulation, our objective is to predict the entire $\mathbf{y}^{(\text{query})} \equiv (\text{T}_{\text{gas}}, \text{p}_{\text{gas}})$ profile for an unknown $\mathbf{x}^{\text{query}}$. 

\begin{figure}[ht!]
\centering
\includegraphics[width=.9\columnwidth]{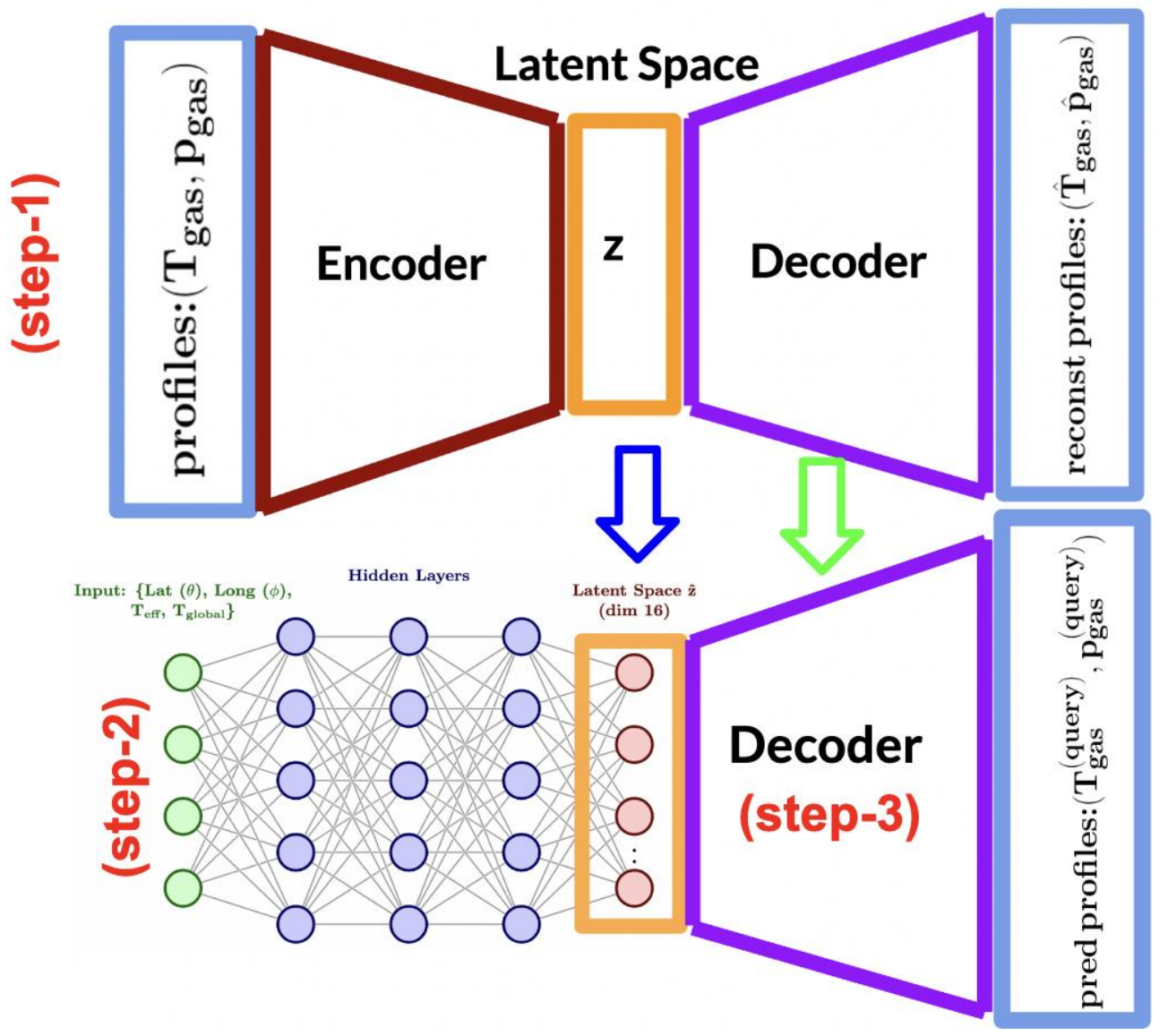}
\caption{\textbf{Step 1} involves designing an optimal AE capable of reconstructing profiles with high accuracy. The training of the AE is constrained based on the loss function defined in Eq. \ref{eq:loss_ae}. In \textbf{step 2}, we train a DNN to learn the relation between input parameters ($\textbf{x}$) and latent space ($\textbf{z}$). The loss function for the training of the DNN model is shown in Eq. \ref{eq:loss_reg}. \textbf{Step 3} performs the inference to obtain a new profile based on a new query point ($\textbf{x}^{\text{query}}$ as shown in Eq. \ref{eq:pred_query}.  }
\label{fig:lsr_framework} 
\end{figure}

\vspace{0.1cm}
\noindent \textbf{(a) Latent Space Regressor:} 
To predict the local gas temperature-pressure ($\text{T}_{\text{gas}}$-$\text{p}_\text{gas}$) profiles from the input parameters $(\text{T}_\text{global}$, $\text{T}_\text{eff}$, $\phi$, $\theta$), we have developed a machine learning framework which has several components. The first component is an autoencoder (AE) architecture \cite{tschannen2018recent} that compresses high-dimensional ($\text{T}_{\text{gas}}$-$\text{p}_{\text{gas}}$) profiles onto a constrained latent manifold. The AE model consists of an asymmetric encoder-decoder architecture in which the input vector—representing local gas-temperature ($\text{T}_{\text{gas}}$) values across discrete pressure layers—is mapped through a series of fully connected layers with decreasing numbers of neurons. Each layer uses a rectified linear unit (ReLU) activation function to capture nonlinearity in the profiles, followed by a bottleneck layer that defines the latent-space representation. The decoder then symmetrically projects these latent vectors back into the original physical dimension. This entire process is referred to as ``reconstruction of the profiles ($\text{T}_{\text{gas}}$-$\text{p}_\text{gas}$)". To ensure a reliable reconstruction, the network is optimized by minimizing the Mean Squared Error (MSE) between the original and reconstructed profiles using an AE. This approach reduces the dimensionality of the original data, which is important if we want to integrate other neural network models as needed. Once the encoder and decoder components are optimized, the corresponding latent space can serve as proxy data to connect the input variables to it. That means instead of designing a multi-layer perceptron which can learn the relation between input parameters and output ($\text{T}_{\text{gas}}$-$\text{p}_\text{gas}$) profiles directly, we can design a simpler architecture (with fewer hidden layers) to learn the relation between input parameters and latent space representation of ($\text{T}_{\text{gas}}$-$\text{p}_\text{gas}$) profiles.  Mapping the latent space with input parameters is the second component of our ML framework. Once we have a trained dense neural network that can predict latent space vectors from the unknown input variables, we need to plug in the trained decoder, which was already optimized during training, to obtain predictions in the original scale. We define this as the third component in the proposed framework. The schematic representation of the framework is shown in Figure \ref{fig:lsr_framework}. 

\noindent \textbf{Mathematical Formulation:} \\
\noindent \textbf{Step-1: Dimensionality Reduction-} First, we need to design an encoder ($\text{E}$), such that the loss function $\mathcal{L}_{\text{AE}}$ can be optimized as follows:
\begin{equation}
\min_{\omega_{1}, \omega_{2}} \mathcal{L}_{\text{AE}} = \frac{1}{N} \sum_{i = 1}^{N} \| \mathbf{y}_i - {\text{D}_\omega}_{1}({\text{E}_\omega}_{2}(\mathbf{y}_i)) \|^2
\label{eq:loss_ae}
\end{equation}
Using encoder, $\text{E}$, we first compress the profiles, hence, $z = \text{E}(y)$ represents the latent space, a mapping from 47 dimension (original resolution of profiles) to a lower dimension dim(z) = 16. 

\noindent \textbf{Step-2: Latent Space Mapping-}
A multi-layer perceptron can be used to map input parameters x with z, and the training would involved a loss function $\mathcal{L}_{\text{reg}}$
\begin{equation}
\min_{\psi} \mathcal{L}_{\text{Reg}} = \frac{1}{N} \sum_{i = 1}^{N} \| \mathbf{z}_i - h_\psi(\mathbf{x}_i) \|^2, \quad \text{where } \mathbf{z}_i = {\text{E}_\omega}_{2}(\mathbf{y}_i)
\label{eq:loss_reg}
\end{equation}

\noindent \textbf{Step-3: Inference/Prediction-} To predict a profile (${\text{T}_{\text{gas}}}^{(\text{query})}, {\text{p}_{\text{gas}}}^{(\text{query})})$ for a new query point $x^{(\text{query})}$:
\begin{equation}
\hat{\mathbf{y}}^{\text{query}} = {\text{D}_\omega}_{1}(h_\psi(\mathbf{x}^{\text{query}}))
\label{eq:pred_query}
\end{equation}
\noindent \textbf{Autoencoder Architecture:} We implement a Symmetric Dense Autoencoder to serve as a baseline for $(\text{T}_{\text{gas}}, \text{p}_{\text{gas}})$ profile compression. This architecture utilizes a feed-forward neural network structured into an encoder and a decoder. The encoder maps the input pressure-layer temperatures through a high-capacity hidden layer of 512 neurons, utilizing a Rectified Linear Unit (ReLU) activation function to introduce non-linearity before compressing the data into the bottleneck latent dimension. The decoder mirrors this structure, projecting the latent representation back through a 512-neuron hidden layer to reconstruct the original input dimensions. This configuration prioritizes a direct, point-to-point mapping of the thermal structure, focusing on capturing the primary variance within the dataset through dense connectivity.

\vspace{0.1cm}
\noindent \textbf{(b) Generative Predictor via CVAE:}
We have explored a CVAE architecture, a simple but effective generative framework that maps high-dimensional (original) profiles to a lower-dimensional latent space while conditioning on other essential parameters. 

\noindent The CVAE architecture comprises three primary components designed for latent representation and profile prediction. 
\begin{itemize}
\item \textbf{Encoder Network ($\text{q}_{\eta}$):} This module receives a concatenated input vector $\mathbf{v}_{\text{in}} = [\mathbf{x}, \mathbf{y} ]$, where $\mathbf{y} \in \mathbb{R}^{47}$ represents the profiles and $\mathbf{x} \in \mathbb{R}^{4}$ represents the input parameters. The information is compressed through a series of fully connected layers: $\mathbb{R}^{51} \rightarrow 128 \rightarrow 64 \rightarrow 32$. 

\item \textbf{Latent Bottleneck:} The final encoder layer maps to two vectors. Instead of a single vector, the encoder outputs two 16-dimensional vectors representing the mean ($\mu$) and log-variance ($\log\sigma^{2} \in \mathbb{R}^{16}$) of the latent distribution. A reparameterization scheme is applied to sample a latent vector ($\mathbf{z} = \mu + \sigma \odot \epsilon$, where $\epsilon \in \mathcal{N}(0, 1)$ ensuring the model remains end-to-end differentiable.
\item \textbf{Decoder Network ($\text{p}_{\lambda}$):}
The decoder mirrors the encoder's structure but functions as a generative mapping. It receives a concatenated vector $[\mathbf{z}, \mathbf{x}] \in \mathbb{R}^{20}$ and passes it through layers: $\mathbb{R}^{20} \rightarrow 64 \rightarrow 128 \rightarrow 47$ to reconstruct the original profile $\hat{\mathbf{y}}$. 

\end{itemize}

\noindent \textbf{Mathematical Formulation:}
CVAE represents the end-to-end probabilistic approach in which the latent space is conditioned on the input parameters $\mathbf{x}$. 

\noindent \textbf{Training Phase:} 
During the training phase, the model learns to generate an output $\textbf{y}$ based on the condition parameters $\textbf{x}$ and the latent variable $\textbf{z}$, which are connected through the following relation:
\begin{equation}
\text{p}(\textbf{y}|\textbf{x}) = \int \text{p}(\textbf{y}|\textbf{z}, \textbf{x}) \, \text{p}(\textbf{z}|\textbf{x}) \, d\textbf{z} \,  
\end{equation}
\noindent The model learns the relationship $\text{p}(\mathbf{y}|\mathbf{x})$ by forcing the latent space $\mathbf{z}$
to capture only the residual information not already provided by the four conditional parameters. By conditioning both the encoder and decoder on $\mathbf{x}$, we ensure the latent space is optimized for variations specific to those parameters.
To train the model, we minimize a loss function composed of two main terms, the reconstruction and the regularization term.
\begin{equation}
\begin{aligned}
\mathcal{L}_{\text{CVAE}}(\lambda, \eta; \mathbf{y}, \mathbf{x}) &= - \mathbb{E}_{q_\eta(\mathbf{z}|\mathbf{y}, \mathbf{x})} \left[ \log p_\lambda(\mathbf{y}|\mathbf{z}, \mathbf{x}) \right] \\
    &\quad + D_{KL} \left( q_\eta(\mathbf{z}|\mathbf{y}, \mathbf{x}) \parallel p_{\lambda}(\mathbf{z}|\mathbf{x}) \right) \, , 
\end{aligned}
\end{equation}
The decoder $p_{\lambda}$ reconstructs $\textbf{y}$ using both the latent representation $\mathbf{z}$ and condition parameter $\mathbf{x}$. 
\noindent The KL Divergence acts as a regularization term, forcing the encoder’s  distribution to ($q_{\eta}$) to match a prior. Crucially, because the prior is also conditioned on, it ``subtracts" the information $\textbf{x}$ provides, ensuring stays independent of the conditioned parameters.
\vspace{0.1cm}

\noindent \textbf{Inference (Prediction) Phase:} Once the model is trained, the encoder is discarded, and the decoder is used as a generative tool. To generate a new output $\mathbf{y}^{(\text{query})}$ corresponding to a specific condition $\mathbf{x}^{(\text{query})}$, we follow two steps:
\begin{enumerate}
\item \textbf{Sampling:} We draw a latent vector $\mathbf{z}$ from the prior distribution $p(\mathbf{z})$, typically a standard Gaussian $\mathcal{N}(\mathbf{0}, \mathbf{I})$. This vector represents the residual variation not captured by the conditions $\mathbf{x}^{(\text{query})}$.
\item \textbf{Generation:} We pass both the sampled $\mathbf{z}$ and the chosen condition $\mathbf{x}^{\text{(query)}}$ into the decoder. The generated output is then sampled from the learned distribution:
\begin{equation}
\mathbf{y}^{\text{(query)}} \sim p_{\lambda}(\mathbf{y} | \mathbf{z}, \mathbf{x}^{\text{(query)}})
\end{equation}

\end{enumerate}
For predicting profiles of unknown atmospheric states, the encoder is bypassed. Given a new set of parameters $\mathbf{x}^{\text{query}}$, we define the latent vector as the distribution mean, $\mathbf{z} = \mathbf{0}$, (or sample for probabilistic estimates). This vector is concatenated with and passed through the decoder to generate the most statistically likely 47-dimensional profile corresponding to those inputs.

\section{Results}
\textbf{Experimental setup:} For this specific study, the scope was narrowed to focus exclusively on \textbf{F-type} host stars. Hence, a comprehensive dataset of 38,800 (= 12 $\times$ 72 $\times$ 45) individual profiles has been considered. These profiles represent equatorial gas-temperature and pressure maps ($\text{T}_{\rm gas}$, $\text{p}_{\rm gas}$) across the 12 global temperature regimes from 400 K to 2600 K. 

\noindent \textbf{Training and Testing Strategy:} To ensure robust training and testing, the entire dataset has been divided into $80\%$ Training and $20\%$ Testing subsets. A stratified random sampling approach was employed across the temperature dimension, as the number of spatial (longitude, latitude) grids is fixed. The total of 38,880 profiles is divided into 12 distinct ``slices," each representing a specific global temperature range from 400K to 2600K. Within each temperature slice, data points are randomly shuffled and split. This ensures that both the training and test sets have equal representation across all temperature levels, preventing the model from becoming biased toward any specific thermal regime. We used the same training and test data with both LSR and CVAE to perform a head-to-head comparison and determine which model has better predictive capacity. 

\noindent \textbf{Statistical Measures:} We evaluated the reconstruction and prediction error for LSR and Generative predictor via CVAE framework using root mean square error (RMSE) and the coefficient of determination ($\text{R}^2$). RMSE measures the average magnitude of the reconstruction error in the same units as the data. $\text{R}^2$ represents the proportion of variance in the profiles captured by LSR and CVAE during reconstruction and prediction. Since the LSR framework involves training of two neural network models, we need to compute the $\text{R}^{2}$ for both the AE and the DNN. In the context of AE, we compute the $\text{R}^{2}$ for the reconstructed profiles of the training dataset. For both LSR (i.e., AE + DNN) and CVAE, we obtain the $\text{R}^{2}$ on the test dataset. In this benchmarking context, an ideal reconstruction is characterized by ($\text{R}^2 \rightarrow 1$) and ($\text{RMSE} \rightarrow 0$). Specifically, 
 $\text{R}^2$ approaching 1 indicates that the model explains nearly all the variability in the original data, while an RMSE approaching 0 signifies that the reconstructed/predicted profiles are nearly identical to the original one, indicating minimal information loss during latent space representation or predicting through DNN or CVAE. 

The selection of \(\text{R}^{2}\) as our primary statistical metric was deliberate. While absolute error metrics like RMSE provide a sense of magnitude, they offer a standardized assessment of the proportion of variance captured by the model. Given that gas-temperature profiles vary significantly in scale, from 400K to 2600K, this allows for a ``temperature-agnostic" comparison. Our results demonstrate that while the LSR model provides a strong baseline, the CVAE consistently achieves a tighter distribution of \(\text{R}^{2}\) scores closer to unity. 

\begin{figure}[ht!]
\centering
\includegraphics[width=\columnwidth]{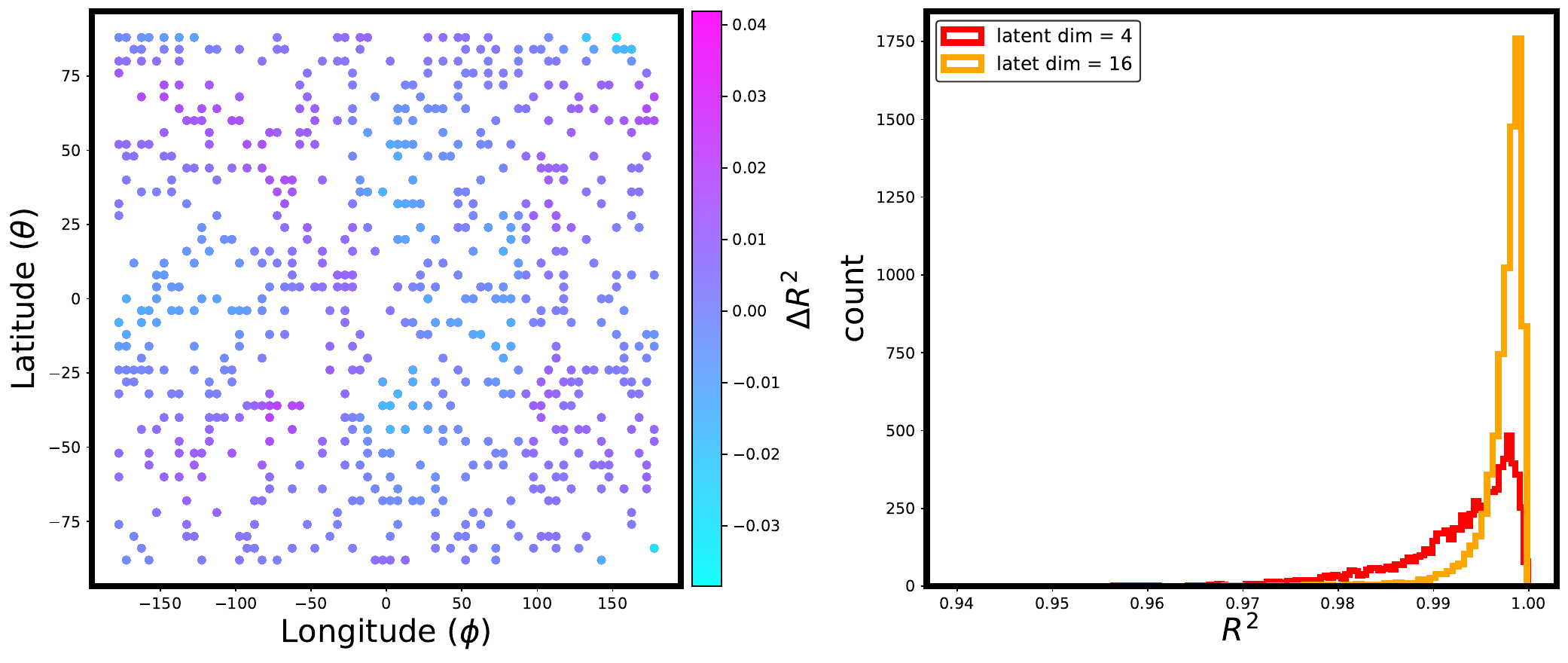}
\caption{\textbf{Left panel:} The prediction accuracy for a set of test grid points (i.e., total 972) is shown using the LSR model. \textbf{Right panel:} Histogram of reconstruction accuracy w.r.to $\Rsq$ for latent dimension 4 and 16 respectively using AE. This exercise has been performed only with a fixed number of training data points corresponding to $\text{T}_{\text{global}} = 2600$ K.}
\label{fig:comp_lsr_4_16} 
\end{figure}

\noindent \textbf{Latent Space Optimization:} To identify the optimal latent space, we explored dimensions of 4, 8, and 16. These three choices are motivated by the goal of compressing the input data by factors ranging from 12$\times$ to 3$\times$. Reconstruction accuracy was evaluated using $\Rsq$ scores for each $\tpprof$. As illustrated in the right panel of Figure \ref{fig:comp_lsr_4_16}, a latent dimension of 16 yields higher $\Rsq$ values, which means better profile reconstruction compared to a dimension of 4. Although not explicitly visualized in the figure, reconstructed profiles from a latent space of dimension 16 are also better than those for the same reconstructed one using dimension 8. This trend is expected because a higher latent dimensionality reduces the severity of the information bottleneck, thereby minimizing feature loss and enhancing reconstruction accuracy. The left panel shows the prediction accuracy of the LSR model. Although this optimization exercise was carried out across all values of $\text{T}_{\text{global}}$, Figure \ref{fig:comp_lsr_4_16} specifically highlights the results for the LSR model trained and tested at $\text{T}_{\text{global}} = 2600$ K. This specific temperature has been selected based on prior empirical observations, which indicated that the reconstruction accuracy of the LSR model inherently degrades at higher temperatures. Hence, it is important to identify at which latent space the LSR model would perform better across the entire $\text{T}_{\text{global}}$ range. While a lower latent dimension 4 may be sufficient for lower temperature ranges, universally enforcing a restrictive bottleneck would compromise prediction accuracy at higher $\text{T}_{\text{global}}$.

\begin{figure}[ht!]
\centering
\includegraphics[width=.95\columnwidth]{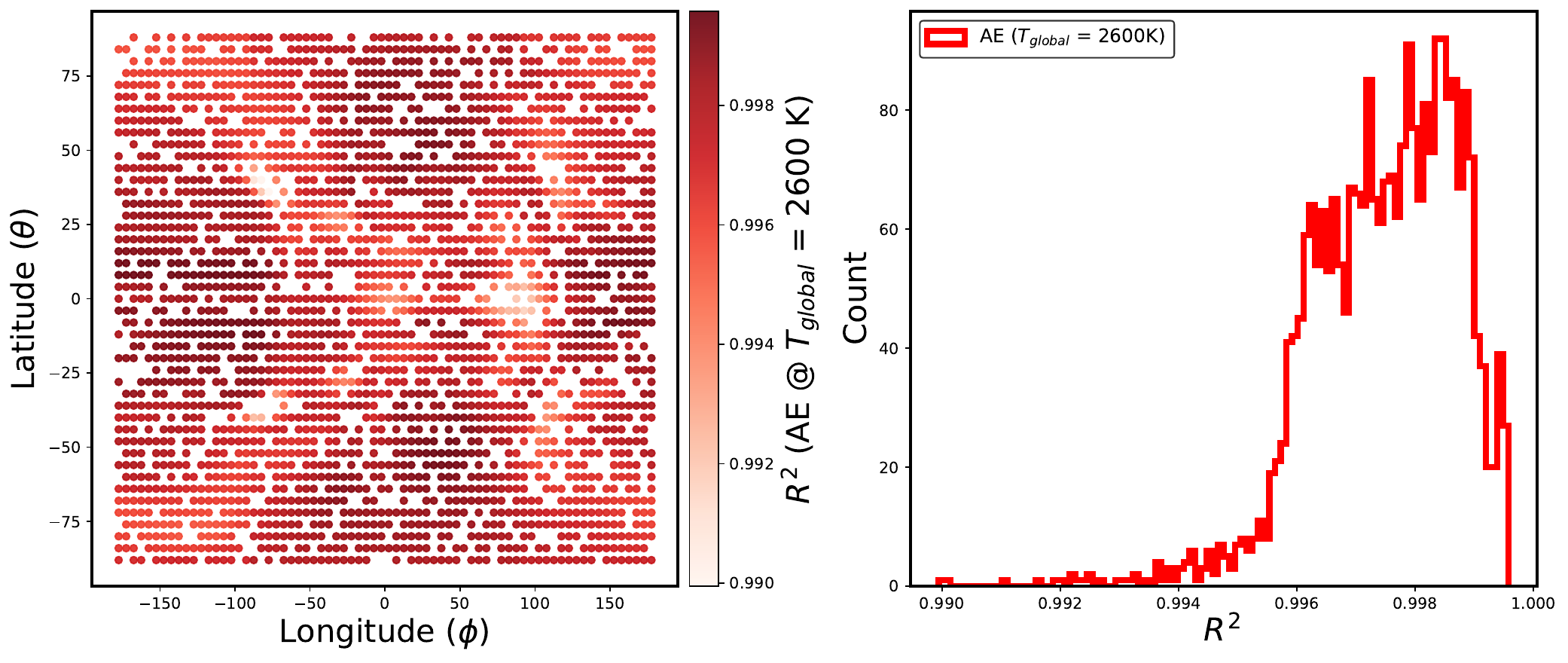}
\caption{\textbf{Left panel:} The left panel shows the reconstruction accuracy of all the $(\text{T}_{\text{gas}}, \text{p}_{\text{gas}})$ profiles w.r.to $\text{T}_{\text{global}} = 2600$ K in terms of $\text{R}^{2}$ for all grid points, represented in $(\phi, \theta)$. \textbf{Right panel:} The right panel shows the histogram of the same $\text{R}^{2}$ values. Since the $\text{R}^{2}$ range is (0.99, 1.0), this indicates that the AE model correctly retrieves these profiles from the latent (embedding) space of dimension 16; hence, the reconstruction accuracy is high. }
\label{fig:ae_train_T_2600} 
\end{figure}

\noindent \textbf{Performance Analysis:} The reconstruction and prediction accuracy for the LSR framework have been shown in Figure \ref{fig:ae_train_T_2600}. It is important to note that, in both reconstruction and prediction cases, the $\text{R}^{2}$ is calculated after transforming the predictions from the normalized scale to the original scale.

\begin{figure}[ht!]
\centering
\includegraphics[width=0.9\columnwidth]{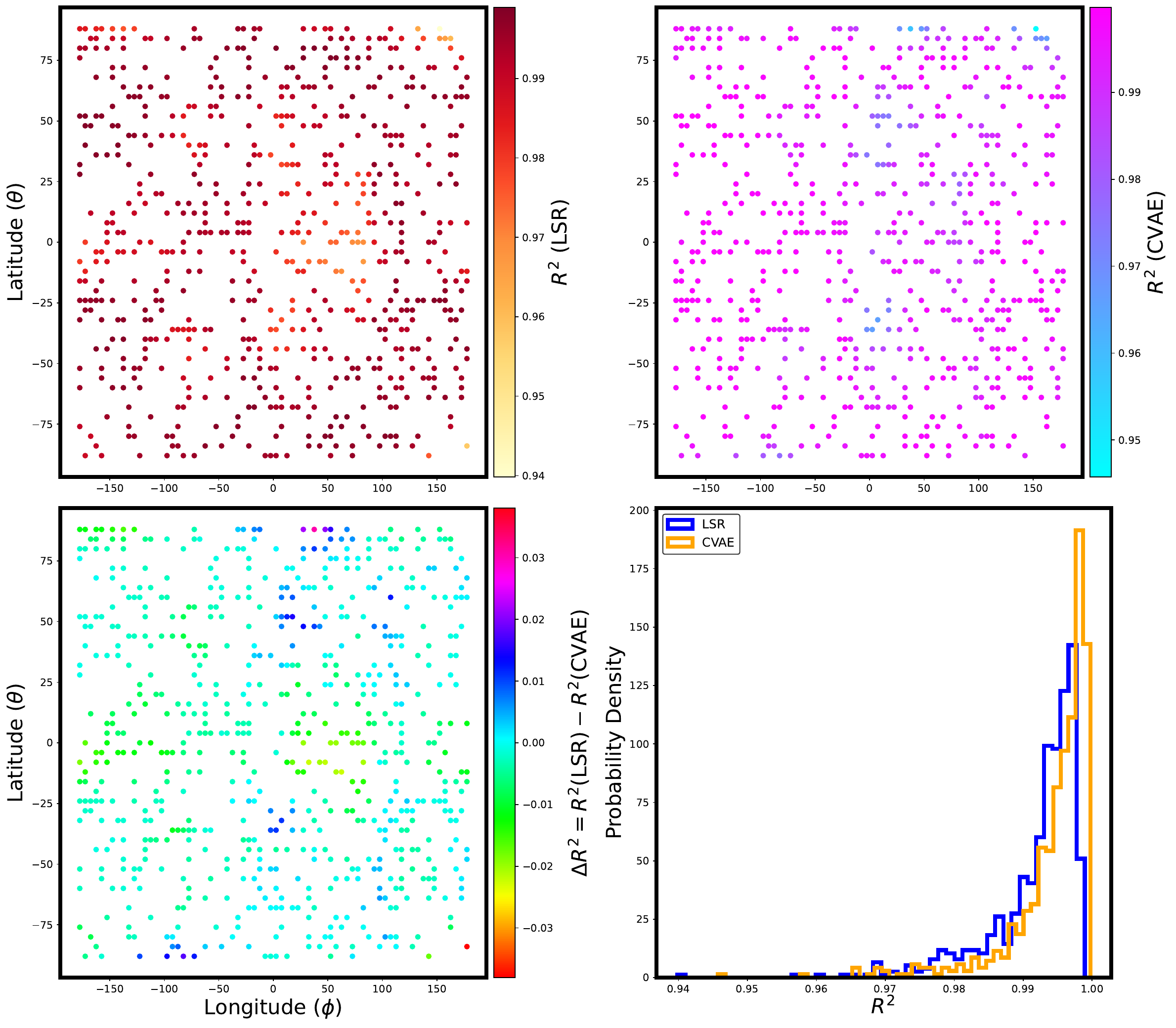}
\caption{\textbf{Left panel:} The left panel shows the difference of $\text{R}^{2}$ obtained from two different method LSR and CVAE for test $(\text{T}_{\text{gas}}$, $\text{p}_{\text{gas}})$ profiles at $T_{\text{global}}$ = 2600 K. \textbf{Right panel:} The right panel shows the probability distribution of $\text{R}^{2}$. }
\label{fig:diff_hist_T_2600} 
\end{figure}

\begin{figure}[ht!]
\centering
\includegraphics[width=0.95\columnwidth]{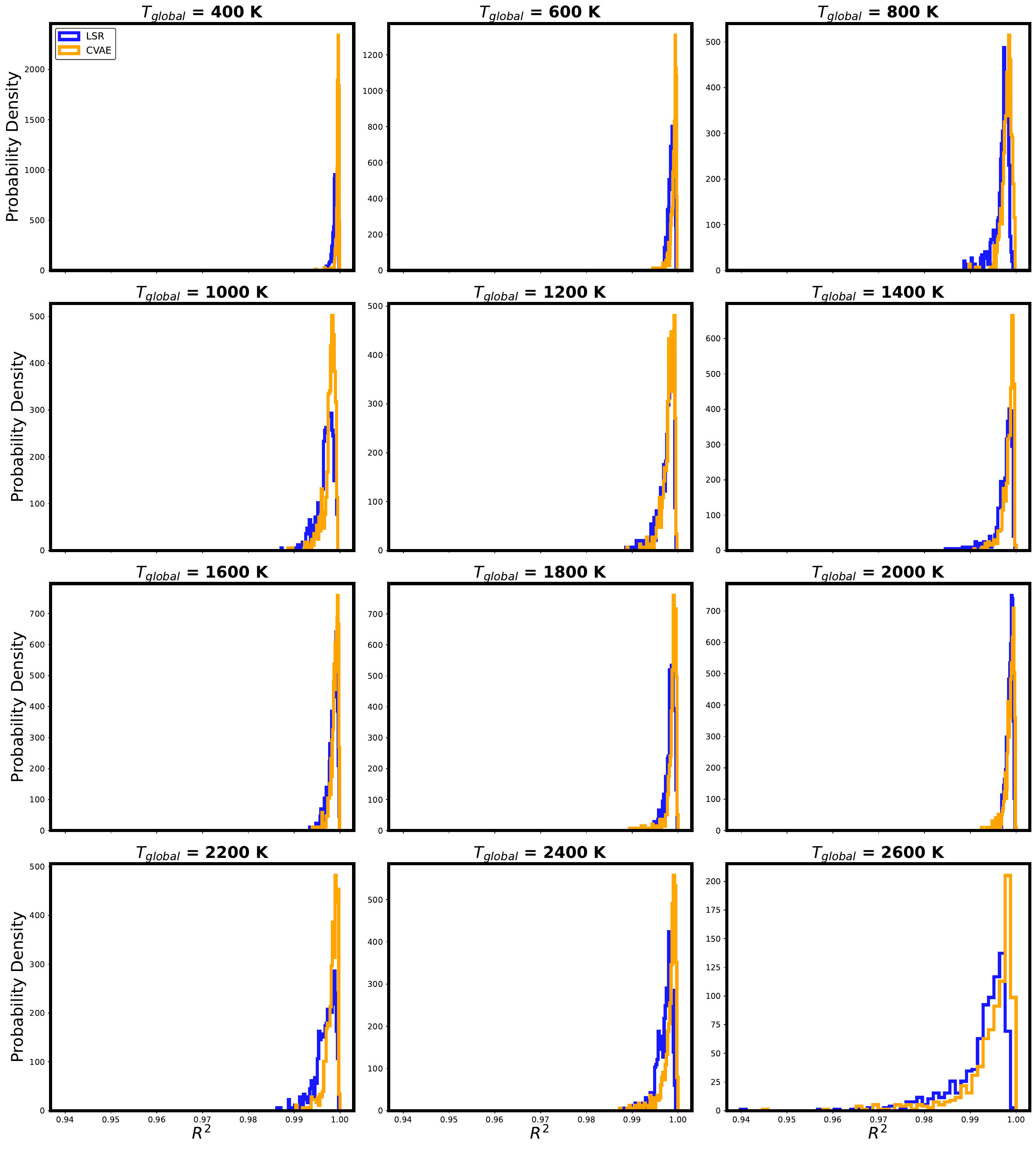}
\caption{The probability density of $\text{R}^{2}$ for all test profiles has been shown and categorized based on their corresponding $\text{T}_{\text{global}}$ values, varying between 400 K to 2600 K. There is a significant overlap between the distribution for both LSR and CVAE. This implies that both methods successfully predict profiles for the test grid points shown in Figure \ref{fig:diff_hist_T_2600}.}
\label{fig:all_T_global_density_comp} 
\end{figure}

\begin{figure}[ht!]
\centering
\includegraphics[width=0.9\columnwidth]{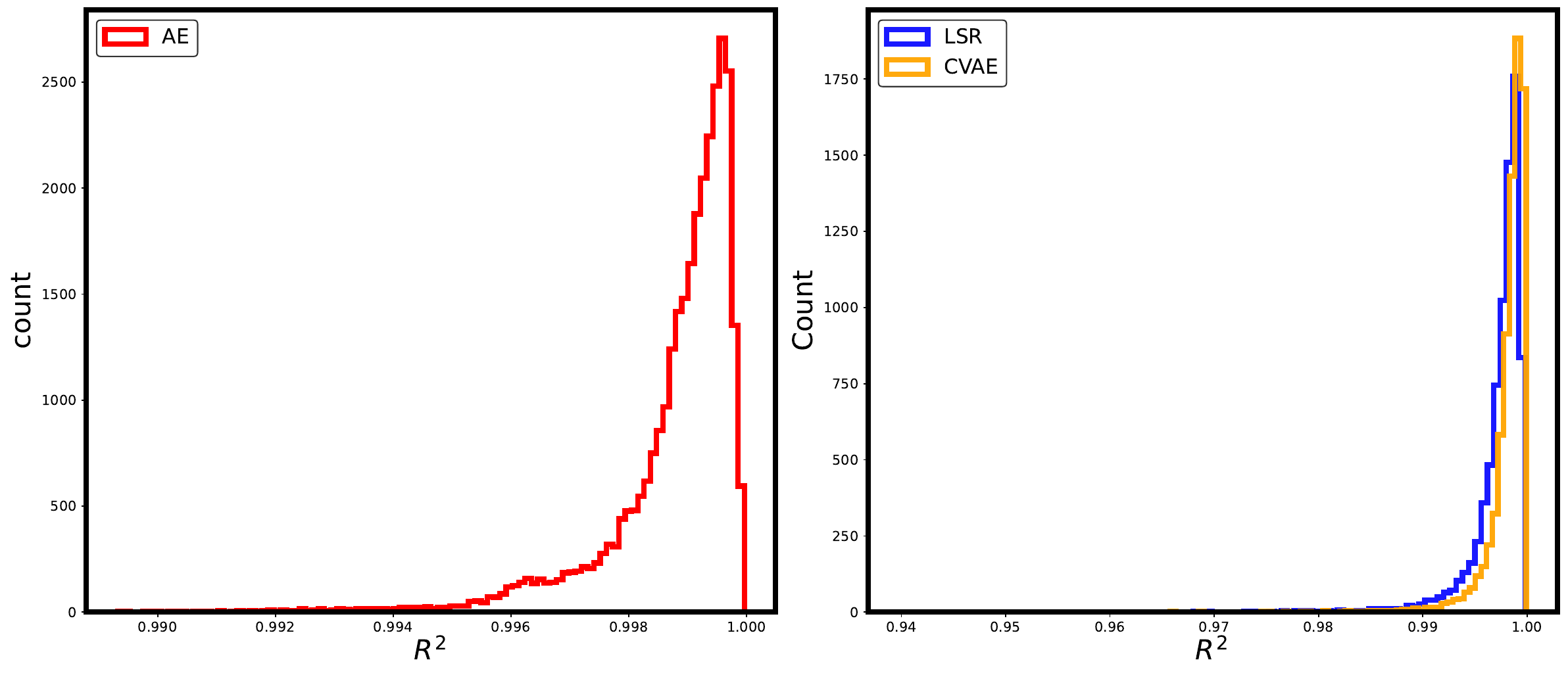}
\caption{\textbf{Left panel:} The left panel shows the histogram of $\text{R}^{2}$ values calculated for the reconstructed profiles using the autoencoder during training of the LSR framework. All the training grids, including the entire $\text{T}_{\text{global}}$ range, are considered. \textbf{Right panel:} The right panel shows the $\text{R}^{2}$ values for the predicted profiles on the unknown test grid for both the LSR and CVAE methods.}
\label{fig:combined_r2} 
\end{figure}

\noindent Figure \ref{fig:diff_hist_T_2600} shows the prediction accuracy for both LSR and CVAE using the exact same test set of 7,760. The similarity of the $\text{R}^{2}$ score distributions indicates that the predictive capability of both models is nearly equal for this specific test dataset. Figure \ref{fig:all_T_global_density_comp} shows a further comparison of the same statistical measure as the global temperature is varied from 400 to 2600. Overall (see Figure \ref{fig:combined_r2}), the CVAE's predictive performance is slightly better than LSR. 

\section{Discussion}
This work presents the first proof of concept that advanced generative models, such as CVAE, can be used to predict local gas-temperature profiles. The proposed framework can be considered a viable alternative to the dense neural network-based model presented in \cite{plaschzug2026accelerating}. By evolving the regression task, the CVAE framework offers a more sophisticated alternative to traditional neural network-based regression schemes. Moreover, unlike standard deep neural networks that rely on pointwise mapping between coordinates and local gas temperatures, the LSR and CVAE architectures learn relationships across the entire profile simultaneously. This method reduces training time and can be efficiently scaled to bigger grids with high point densities. Furthermore, treating profiles as samples from a learned conditional distribution equips the CVAE to better capture complex physical phenomena. While current results are demonstrated for an F-type star, both methods are capable of predicting profiles for other host-star types with high accuracy. Future work will involve a detailed analysis of these models across diverse stellar types and planetary environments highly relevant to upcoming PLATO observations.

\begin{acknowledgments}
The authors gratefully acknowledge Alexander Plaschzug and Sebastian Gernjak for their foundational work during their Master's theses. Their contributions to the initial dense neural network and decision-tree-based emulators laid the groundwork for the follow-up work on the conditional variational autoencoder (CVAE) and latent space regressor (LSR) extensions presented here. We also thank the two anonymous reviewers for their constructive comments and feedback that helped to improve the quality of this manuscript. Further, we would like to acknowledge the Vienna Science Cluster (VSC) HPC facility for providing CPU time for the 3D GCMs for Exoplanets project (72245).
\end{acknowledgments}
\printbibliography
\addcontentsline{toc}{section}{References}

\end{document}